\documentclass[fp,dvipdfmx]{jpsj3}
\usepackage[dvipdfmx]{graphicx}
\usepackage[dvipdfmx]{color}
\renewcommand{\a}{\alpha}
\renewcommand{\b}{\beta}
\newcommand{\bea}{\begin{eqnarray}}
\newcommand{\eea}{\end{eqnarray}}
\newcommand{\f}[2]{\frac{#1}{#2}}
\newcommand{\eq}{&=&}
\newcommand{\nn}{\nonumber \\ }
\newcommand{\ve}{\varepsilon}

\newcommand{\area}{\int_{-\infty}^\infty }

\newcommand{\p}{\partial}
\newcommand{\pp}[2]{\f{\p #1}{\p #2}}

\newcommand{\siki}[1]{Eq. (\ref{#1})}
\newcommand{\sikis}[2]{Eqs. (\ref{#1}) and (\ref{#2})}

\allowdisplaybreaks[1]

\allowdisplaybreaks[1]

\title{
Minimal Investment Risk\\
with Cost and Return Constraints: A Replica Analysis
}

\author{Takashi Shinzato\thanks{shinzato@eng.tamagawa.ac.jp}
}
\inst{
Department of Management Science, College of Engineering, 
Tamagawa University, \\ Machida, Tokyo 1948610, Japan\\
\smallskip
{\rm (Received January 30, 2019; accepted *** 1, 2019)}
} 
\abst{
Previous studies into the budget constraint of portfolio optimization problems based on statistical mechanical informatics have not considered that the purchase cost per unit of each asset is distinct. Moreover, the fact that the optimal investment allocation differs depending on the size of investable funds has also been neglected. In this paper, we approach the problem of investment risk minimization using replica analysis. This problem imposes cost and return constraints. We also derive the macroscopic theory indicated by the optimal solution and confirm the validity of our proposed method through numerical experiments.
}

\begin{document}
\maketitle

\section{Introduction}
Extant literature in the domain of operations research has analyzed annealed disordered systems in the context of spin glass theory against portfolio optimization problems such as budget constrained investment risk minimization problems and risk constrained expected return maximization\cite{Luenberger1998InvestmentScience,marcus2014investments}. However, the investment information sought by investors is actually the optimal portfolio in the quenched disordered system of the investment market. Thus, in recent years, researchers have actively analyzed these portfolio optimization problems using statistical mechanical informatics represented by random matrix theory, replica analysis, and the belief propagation method\cite{
1742-5468-2017-12-123402,
1742-5468-2016-12-123404,
doi:10.1080/1351847X.2011.601661,
KONDOR20071545,
PAFKA2003487,
Pafka2002,
Ciliberti2007,
doi:10.1080/14697680701422089,
110008689817,
doi:10.7566/JPSJ.86.063802,
doi:10.7566/JPSJ.86.124804,
SHINZATO2018986,
PhysRevE.94.052307,
PhysRevE.94.062102b,
10.1371/journal.pone.0134968,
10.1371/journal.pone.0133846,
1742-5468-2017-2-023301,
2018arXiv181006366S,
doi:10.7566/JPSJ.87.064801,
1742-5468-2018-2-023401,
Ryosuke-Wakai2014}. Through these studies, it is possible to analyze the quenched disordered system of the investment market, which was hitherto difficult to analyze by applying the well-used analysis methods of operations research. These studies in cross-disciplinary research fields also 
could analyze the mathematical structure of the minimum investment risk, the concentrated investment, and the maximum expected return\cite{
110008689817,
doi:10.7566/JPSJ.86.063802,
doi:10.7566/JPSJ.86.124804,
SHINZATO2018986,
PhysRevE.94.052307,
PhysRevE.94.062102b,
10.1371/journal.pone.0134968,
10.1371/journal.pone.0133846,
1742-5468-2017-2-023301,
2018arXiv181006366S,
doi:10.7566/JPSJ.87.064801,
1742-5468-2018-2-023401,
Ryosuke-Wakai2014}.

However, although the budget constraint is used as a representative constraint condition in portfolio optimization problems approached using statistical mechanical informatics,  we impose the strong assumption that purchase costs per unit of each asset are the same for all assets. 
Furthermore, previous studies used problem settings relevant to operations research; thus, we focus on the investment ratio of each asset as a decision variable, regardless of investment fund size. However, due to the size of working capital in actual investment contexts, the optimal investment strategy of individual investors with low working capital and the optimal investment strategy of institutional investors with sufficiently large working capital are different.

Therefore, in the present paper, we improve on the analytical approaches of previous works that utilized statistical mechanical informatics and discuss the investment risk minimization problem imposing cost and return constraints by using replica analysis. We also derive the macroscopic theory satisfied by the optimal portfolio.

The remainder of the paper is organized as follows. The next {two sections describe} the model setting {and} the replica analysis used to solve the portfolio optimization problem imposing constraints of initial cost and final return. 
Section \ref{sec4} discusses the optimal portfolio in several situations and the macroscopic relations of the optimal solution. Numerical experiments confirm the validity of our proposed method based on replica analysis. The final section offers a  summary and discusses potential future work in this domain.

\section{Model setting}
Let us consider a situation whereby $N$ assets are invested for $p$ periods in a steady trading market. {Similar to the related literature by using replica analysis}, we assume that no short selling regulation is imposed on the investment market.
We denote the portfolio of asset $i(=1,2\cdots,N)$ as $w_i\in{\bf R}$,
such that the vector $\vec{w}=(w_1,w_2,\cdots,w_N)^{\rm T}\in{\bf R}^N$
describes the portfolio of $N$ assets, where the notation ${\rm T}$ represents the transpose of the vector and/or matrix. Moreover, the purchase cost per unit of asset $i$ at the initial investment period is expressed by 
$c_i$ and 
the return per unit of asset $i$ at period $\mu(=1,2,\cdots,p)$
is represented by $\bar{x}_{i\mu}$. We also assume that each return is independently distributed and the mean $E[\bar{x}_{i\mu}]=r_i$ and 
variance of the return 
$V[\bar{x}_{i\mu}]=v_i$
are known. Next, 
we assume that portfolio 
$\vec{w}$ imposes 
the cost constraint in \siki{eq1}
and the return constraint in \siki{eq2};
\bea
\label{eq1}
N\times C\eq\sum_{i=1}^Nc_iw_i,\\
\label{eq2}
N\times R\eq
\sum_{i=1}^Nr_iw_i,
\eea
where $NC$ is the initial budget at the initial investment period
and $NR$ is the final return at the last investment period.  

The coefficients $C$ and $R$ denote the unit cost per asset and the unit return per asset, respectively. In practice, since 
the purchase costs per unit of each asset 
$c_i$ do not always coincide,
in this paper, we do not apply the budget constraint used in previous studies 
$\sum_{i=1}^Nw_i=N$, but we apply the cost constraint in \siki{eq1}. Thus, 
the feasible subspace of portfolio $\vec{w}$,
${\cal W}\subseteq{\bf R}^N$, is defined by
\bea
{\cal W}\eq
\left\{\vec{w}\in{\bf R}^N\left|
NC=\vec{w}^{\rm T}\vec{c},
NR=\vec{w}^{\rm T}\vec{r}
\right.\right\},
\eea
where  
$\vec{c}=(c_1,c_2,\cdots,c_N)^{\rm T}\in{\bf R}^N$
 and $\vec{r}=(r_1,r_2,\cdots,r_N)^{\rm T}\in{\bf R}^N$ are used.

From this, the investment risk of portfolio $\vec{w}$, 
${\cal H}(\vec{w})$, is as follows:
\bea
\label{eq4}
{\cal H}(\vec{w})\eq
\f{1}{2N}\sum_{\mu=1}^p
\left(
\sum_{i=1}^N\bar{x}_{i\mu}w_{i}-
\sum_{i=1}^Nr_iw_{i}
\right)^2\nn
\eq
\f{1}{2}\vec{w}^{\rm T}J
\vec{w},
\eea
The $i,j$th component of Wishart matrix $J=\left\{J_{ij}\right\}\in{\bf R}^{N\times N}$,
$J_{ij}$, is given by
\bea
\label{eq5}
J_{ij}\eq\f{1}{N}
\sum_{\mu=1}^p
\left(\bar{x}_{i\mu}-r_i\right)
\left(\bar{x}_{j\mu}-r_j\right)\nn
\eq\f{1}{N}\sum_{\mu=1}^px_{i\mu}x_{j\mu},
\eea
where in \siki{eq5}
the modified return $x_{i\mu}=\bar{x}_{i\mu}-r_i$ is already used, 
its mean and variance are $E[x_{i\mu}]=0$ and 
$V[x_{i\mu}]=v_i$, respectively. Thus, 
the optimal portfolio of the portfolio optimization problem that we discuss $\vec{w}^*$ is described as 
\bea
\vec{w}^*\eq\arg\mathop{\min}_{\vec{w}\in{\cal W}}{\cal H}(\vec{w}).
\eea
We accept $p>N$ herein since 
the optimum can be uniquely determined.

This portfolio optimization problem 
can be solved by using the extremum of the following Lagrange multiplier function ${\cal L}$:
\bea
{\cal L}
=\f{1}{2}\vec{w}^{\rm T}J\vec{w}
+\theta(NR-\vec{w}^{\rm T}\vec{r})
+k(NC-\vec{w}^{\rm T}\vec{c}).
\eea
That is, from the extremum of ${\cal L}$, $\pp{\cal L}{w_i}
=\pp{\cal L}{k}=\pp{\cal L}{\theta}=0
$, the optimal $\vec{w}^*=\arg
\mathop{\min}_{\vec{w}\in{\cal W}}{\cal H}(\vec{w})
$ is derived. Then, the minimal investment risk per asset $\ve=\f{1}{N}{\cal H}(\vec{w}^*)$ is obtained:
\bea
\label{eq8}
\ve
\eq
\f{N}{2}\f{R^2\vec{c}^{\rm T}J^{-1}\vec{c}
-2RC\vec{c}^{\rm T}J^{-1}\vec{r}
+C^2
\vec{r}^{\rm T}J^{-1}\vec{r}
}{\vec{c}^{\rm T}J^{-1}\vec{c}
\vec{r}^{\rm T}J^{-1}\vec{r}
-(\vec{c}^{\rm T}J^{-1}\vec{r})^2
}\nn
\eq
\f{N}{2\vec{c}^{\rm T}J^{-1}\vec{c}}
\left\{
C^2+\f{\left(R-C\f{\vec{c}^{\rm T}J^{-1}\vec{r}}{\vec{c}^{\rm T}J^{-1}\vec{c}}\right)^2}
{\f{\vec{r}^{\rm T}J^{-1}\vec{r}}{\vec{c}^{\rm T}J^{-1}\vec{c}}-\left(\f{\vec{c}^{\rm T}J^{-1}\vec{r}}{\vec{c}^{\rm T}J^{-1}\vec{c}}\right)^2}
\right\},\qquad
\eea
where Eqs. (\ref{eq9})--(\ref{eq11}) are used:
\bea
\label{eq9}
k^*\eq
\f{-NR\vec{r}^{\rm T}J^{-1}\vec{c}
+NC\vec{r}^{\rm T}J^{-1}\vec{r}}
{\vec{c}^{\rm T}J^{-1}\vec{c}
\vec{r}^{\rm T}J^{-1}\vec{r}
-(\vec{c}^{\rm T}J^{-1}\vec{r})^2
},\\
\theta^*\eq
\f{NR\vec{c}^{\rm T}J^{-1}\vec{c}
-NC\vec{c}^{\rm T}J^{-1}\vec{r}}
{\vec{c}^{\rm T}J^{-1}\vec{c}
\vec{r}^{\rm T}J^{-1}\vec{r}
-(\vec{c}^{\rm T}J^{-1}\vec{r})^2
},\\
\label{eq11}
\vec{w}^*\eq
\theta^*J^{-1}\vec{r}+
k^*J^{-1}\vec{c}.
\eea
It transpires that the optimal portfolio $\vec{w}^*$ is dependent on the initial cost $C$ and the final return $R$ from Eqs. (\ref{eq9})--(\ref{eq11}). It is also the case that 
the optimal investment strategy is a function of the size of working capital and the target figure.
If we can assess $\vec{c}^{\rm T}J^{-1}\vec{c},
\vec{c}^{\rm T}J^{-1}\vec{r},
\vec{r}^{\rm T}J^{-1}\vec{r}
$, using \siki{eq8}, 
the minimal investment risk per asset $\ve$ is calculated. 
However, in general, it is computationally onerous to solve for the inverse matrix $J^{-1}$ of the regular matrix $J\in{\bf R}^{N\times N}$ as $N$ increases.
Therefore, we discuss the portfolio optimization problem using replica analysis which can resolve the minimal investment risk per asset $\ve$
without directly solving for the inverse matrix $J^{-1}$.

\section{Replica analysis\label{sec3}}

Following an analytical procedure based on statistical mechanical informatics, 
we discuss an optimization problem that has 
a Hamiltonian of the investment system defined in 
\siki{eq4}. Then 
the partition function $Z$ 
of the inverse temperature 
$\b(>0)$ of the canonical ensemble is defined as 
\bea
Z\eq
\int_{\vec{w}\in{\cal W}} d\vec{w}e^{-\b{\cal H}(\vec{w})}\nn
\eq
\f{1}{(2\pi)^{\f{N}{2}}}
\mathop{\rm Extr}_{k,\theta}
\area 
d\vec{w}
\exp
\left(-\f{\b}{2}\sum_{i=1}^N\sum_{j=1}^Nw_iw_jJ_{ij}
\right.\nn
&&
\left.
+k\left(\sum_{i=1}^Nc_iw_i-NC\right)
+\theta\left(\sum_{i=1}^Nr_iw_i-NR\right)
\right),\nn
\eea
where $k,\theta$ are the variables related to the constraints in \sikis{eq1}{eq2}. From this, 
the minimal investment risk per asset $\ve$ is solved from the following thermodynamic relation: 
\bea
\label{eq13}
\ve\eq-\lim_{\b\to\infty}\pp{\phi}{\b},
\eea
where it is well-known that the minimal investment risk per asset $\ve$
holds if the following self-averaging property is used: 
\bea
\phi\eq
\lim_{N\to\infty}
\f{1}{N}E[\log Z].
\eea
In general, it is cumbersome to directly evaluate the configuration average of 
$\log Z$ over return matrix $X$, $E[\log Z]$. Since it is comparatively easy to execute 
$E[Z^n],n\in{\bf Z}$ using replica analysis in the limit that the number of assets $N$ is sufficiently large,
\bea
\psi(n)\eq
\lim_{N\to\infty}\f{1}{N}\log E[Z^n]
\nn
\eq\mathop{\rm Extr}_{\vec{\theta
},\vec{k},Q_s,\tilde{Q}_s}
\left\{
-\f{\a}{2}\log
\det\left|I+\b Q_s\right|
+\f{1}{2}{\rm Tr}
Q_s\tilde{Q}_s
\right.
\nn
&&-R\vec{\theta}^{\rm T}\vec{e}
-C\vec{k}^{\rm T}\vec{e}-\f{1}{2}\log\det\left|\tilde{Q}_s\right|
-\f{n}{2}
\left\langle
\log v
\right\rangle
\nn
&&
\left.
+\f{1}{2}
\left\langle
\f{1}{v}
(c\vec{k}+r\vec{\theta})^{\rm T}
\tilde{Q}_s^{-1}
(c\vec{k}+r\vec{\theta})
\right\rangle
\right\}
\eea
is analytically evaluated where the period ratio $\a=p/N\sim O(1)$, the order parameters $\vec{\theta}=(\theta_1,\theta_2,\cdots,\theta_n)^{\rm T}\in{\bf R}^n$,
$\vec{k}=(k_1,k_2,\cdots,k_n)^{\rm T}\in{\bf R}^n$,
$Q_s=\left\{q_{sab}\right\}\in{\bf R}^{n\times n}$
$\tilde{Q}_s=\left\{\tilde{q}_{sab}\right\}\in{\bf R}^{n\times n}$, the identity matrix $I\in{\bf R}^{n\times n}$, and constant vector $\vec{e}=(1,1,\cdots,1)^{\rm T}\in{\bf R}^n$
are already used. Moreover, the notation 
\bea
\left\langle
f(r,c,v)
\right\rangle
\eq\lim_{N\to\infty}\f{1}{N}
\sum_{i=1}^Nf(r_i,c_i,v_i),
\eea
is employed. Further, the notation ${\rm Extr}_z g(z)$ denotes the extremum of $g(z)$ by $z$,
and $\tilde{q}_{sab}$ is the auxiliary order parameter of 
\bea
q_{sab}\eq
\lim_{N\to\infty}
\f{1}{N}
\sum_{i=1}^Nv_iw_{ia}w_{ib}.
\eea

Here we assume the replica symmetry solution. Then, 
$\theta_a=\theta,k_a=k$, 
$q_{saa}=\chi_s+q_s$, ${q}_{sab}=q_s$,
$\tilde{q}_{saa}=\tilde{\chi}_s-\tilde{q}_s$, $\tilde{q}_{sab}=-\tilde{q}_s,(a\ne b)$
are set; thus, 
\bea
\psi(n)\eq-\f{\a(n-1)}{2}\log(1+\b\chi_s)-\f{n}{2}
\left\langle
\log v
\right\rangle\nn
&&
-\f{\a}{2}\log(1+\b\chi_s+n\b q_s)-nR\theta-nCk\nn
&&+\f{n}{2}(\chi_s+q_s)(\tilde{\chi}_s-\tilde{q}_s)
-\f{n(n-1)}{2}q_s\tilde{q}_s\nn
&&
-\f{n-1}{2}\log\tilde{\chi}_s-\f{1}{2}\log(\tilde{\chi}_s-n\tilde{q}_s)\nn
&&+\f{n}{2(\tilde{\chi}_s-n\tilde{q}_s)}\left\langle
\f{(ck+r\theta)^2}{v}
\right\rangle,
\eea
is replaced where ${\rm Extr}$ is abbreviated. From this, $\phi
=\lim_{n\to0}\pp{\psi(n)}{n}
$ is summarized as follows:
\bea
\phi
\eq-\f{\a}{2}\log(1+\b\chi_s)-\f{\a\b q_s}{2(1+\b\chi_s)}
-R\theta-Ck
\nn
&&+\f{1}{2}(\chi_s+q_s)(\tilde{\chi}_s-\tilde{q}_s)
+\f{1}{2}q_s\tilde{q}_s
-\f{1}{2}\log\tilde{\chi}_s\nn
&&+\f{\tilde{q}_s}{2\tilde{\chi}_s}
+\f{1}{2\tilde{\chi}_s}\left\langle
\f{(ck+r\theta)^2}{v}
\right\rangle-\f{1}{2}
\left\langle
\log v
\right\rangle.
\label{eq12}
\eea
Moreover, from the extremum condition of \siki{eq12},
\bea
\chi_s\eq\f{1}{\b(\a-1)},\\
q_s\eq\f{\a}{\a-1}\f{R^2\left\langle 
\f{c^2}{v}\right\rangle -2RC
\left\langle
\f{rc}{v}
\right\rangle
+C^2\left\langle 
\f{r^2}{v}\right\rangle
}{
\left\langle \f{r^2}{v}\right\rangle
\left\langle \f{c^2}{v}\right\rangle
-\left\langle\f{rc}{v}\right\rangle^2
},\\
\tilde{\chi}_s\eq\b(\a-1),\\
\tilde{q}_s\eq\b^2(\a-1)
\f{R^2\left\langle 
\f{c^2}{v}\right\rangle -2RC
\left\langle
\f{rc}{v}
\right\rangle
+C^2\left\langle 
\f{r^2}{v}\right\rangle
}{
\left\langle \f{r^2}{v}\right\rangle
\left\langle \f{c^2}{v}\right\rangle
-\left\langle\f{rc}{v}\right\rangle^2
},\nn
\\
\theta\eq\b(\a-1)
\f{R\left\langle \f{c^2}{v}\right\rangle -C
\left\langle
\f{rc}{v}
\right\rangle
}{
\left\langle \f{r^2}{v}\right\rangle
\left\langle \f{c^2}{v}\right\rangle
-\left\langle \f{rc}{v}\right\rangle^2
},\\
k\eq\b(\a-1)
\f{-R
\left\langle
\f{rc}{v}
\right\rangle
+C\left\langle \f{r^2}{v}\right\rangle
}{
\left\langle \f{r^2}{v}\right\rangle
\left\langle \f{c^2}{v}\right\rangle
-\left\langle \f{rc}{v}\right\rangle^2
}
\eea
are obtained. From these results, using the identical equation in \siki{eq13}, $
\ve=-
\lim_{\b\to\infty}\pp{\phi}{\b}$, the minimal investment risk per asset 
$\ve$ is summarized as 
\bea
\label{eq25}
\ve
\eq\f{\a-1}{2\left\langle v^{-1}c^2\right\rangle}
\left(
C^2+\f{\left(R-R_0
\right)^2}{V}
\right),
\eea
from $\pp{\phi}{\b}=-\f{\a\chi_s}{2(1+\b\chi_s)}-\f{\a q_s}{2(1+\b\chi_s)^2}$, where 
\bea
R_0\eq
C
\f{\left\langle v^{-1}rc\right\rangle}
{\left\langle v^{-1}c^2\right\rangle},\\
V\eq
\f{\left\langle v^{-1}r^2\right\rangle}
{\left\langle v^{-1}c^2\right\rangle}
-\left(
\f{\left\langle v^{-1}rc\right\rangle}
{\left\langle v^{-1}c^2\right\rangle}
\right)^2
\eea
are used.
\section{Discussion\label{sec4}}
In the case where the only portfolio constraint concerns cost, $\phi=\lim_{N\to\infty}\f{1}{N}E[\log Z]$ is
\bea
\phi
\eq-\f{\a}{2}\log(1+\b\chi_s)-\f{\a\b q_s}{2(1+\b\chi_s)}
-Ck
\nn
&&+\f{1}{2}(\chi_s+q_s)(\tilde{\chi}_s-\tilde{q}_s)
+\f{1}{2}q_s\tilde{q}_s
-\f{1}{2}\log\tilde{\chi}_s\nn
&&+\f{\tilde{q}_s}{2\tilde{\chi}_s}
+\f{k^2}{2\tilde{\chi}_s}\left\langle
v^{-1}c^2
\right\rangle-\f{1}{2}
\left\langle
\log v
\right\rangle;
\eea
then, the minimal investment risk per asset 
$\ve=-\lim_{\b\to\infty}\pp{\phi}{\b}$ is
\bea
\ve\eq\f{\a-1}{2\left\langle v^{-1}c^2\right\rangle}C^2.
\label{eq46}
\eea
When $C=c_i=1$, the result already available in the literature $\f{\a-1}{2\left\langle v^{-1}\right\rangle}$ is derived\cite{PhysRevE.94.062102b}. Moreover, to compare \sikis{eq25}{eq46},
when the return coefficient $R=R_0$, that is, when the weighted average of the revenue growth rate of asset $i$, $\f{r_i}{c_i}$ is equal to the revenue growth rate of the portfolio $\f{R}{C}$,
$\f{R}{C}=\f{\left\langle v^{-1}c^2\f{r}{c}\right\rangle}{\left\langle v^{-1}c^2\right\rangle}$, it is possible to resolve the minimal investment risk per asset under the cost constraint.

Next,  let us compare 
results according to replica analysis and the  
Lagrange multiplier method that solves 
three moments $\f{1}{N}\vec{r}^{\rm T}J^{-1}\vec{r}$, 
$\f{1}{N}\vec{r}^{\rm T}J^{-1}\vec{c}$, 
$\f{1}{N}\vec{c}^{\rm T}J^{-1}\vec{c}$.
We consider the following partition function:
\bea
Z\eq
\f{1}{(2\pi)^{\f{N}{2}}}
\area d\vec{w}
e^{-\f{1}{2}\vec{w}^{\rm T}J\vec{w}+\theta\vec{w}^{\rm T}\vec{r}+k\vec{w}^{\rm T}\vec{c}}.
\eea
It is straightforward to calculate the integral of the partition function,
\bea
\log Z\eq
-\f{1}{2}\log\det|J|
+\f{\theta^2}{2}\vec{r}^{\rm T}J^{-1}\vec{r}
+\f{k^2}{2}\vec{c}^{\rm T}J^{-1}\vec{c}
\nn
&&+\theta k\vec{r}^{\rm T}J^{-1}\vec{c}.
\eea
Furthermore, 
$\log Z$ {holds} the self-averaging property, 
$\phi=\lim_{N\to\infty}\f{1}{N}E[\log Z]$ is solved, and from the derivative function with respect to $\theta,k$, we can solve the three moments. From the assumption of the replica symmetry solution,
\bea
\phi
\eq-\f{\a}{2}\log(1+\chi_s)-\f{\a q_s}{2(1+\chi_s)}-\f{1}{2}
\left\langle
\log v
\right\rangle\nn
&&+
\f{1}{2}(\chi_s+q_s)(\tilde{\chi}_s-\tilde{q}_s)+\f{1}{2}q_s\tilde{q}_s\nn
&&-\f{1}{2}\log\tilde{\chi}_s+\f{\tilde{q}_s}{2\tilde{\chi}_s}
+\f{\left\langle v^{-1}(ck+r\theta )^2\right\rangle}{2\tilde{\chi}_s}\label{eq32}
\eea
is summarized. Thus, from the extrema, 
$\chi_s=\f{1}{\a-1}$,
$q_s=\f{\a}{(\a-1)^3}\left\langle v^{-1}(ck+r\theta)^2\right\rangle$, 
$\tilde{\chi}_s=\a-1$,
$\tilde{q}_s=\f{1}{\a-1}\left\langle v^{-1}(ck+r\theta)^2\right\rangle$ are obtained. 
Substituting these into \siki{eq32},
\bea
\phi\eq-\f{\a}{2}\log\a+\f{\a-1}{2}\log(\a-1)\nn
&&+\f{1}{2}
+\f{\left\langle v^{-1}(ck+r\theta)^2\right\rangle}{2(\a-1)}-\f{1}{2}
\left\langle
\log v
\right\rangle
\eea
is derived. From this, 
\bea
\lim_{N\to\infty}\f{1}{N}
\vec{r}^{\rm T}J^{-1}\vec{r}
\eq\pp{^2\phi}{\theta^2}\nn
\eq\f{\left\langle v^{-1}r^2\right\rangle}{\a-1},\\
\lim_{N\to\infty}\f{1}{N}
\vec{r}^{\rm T}J^{-1}\vec{c}
\eq\pp{^2\phi}{\theta\p k}\nn
\eq\f{\left\langle v^{-1}rc\right\rangle}{\a-1},\\
\lim_{N\to\infty}\f{1}{N}
\vec{c}^{\rm T}J^{-1}\vec{c}
\eq\pp{^2\phi}{k^2}\nn
\eq\f{\left\langle v^{-1}c^2\right\rangle}{\a-1}
\eea
are obtained. We substitute these into \siki{eq8}; then 
it transpires that this 
is consistent with 
the result in \siki{eq25}.

The Sharpe ratio, which is defined by the return per unit risk, $S=S(R)$ is given by
\bea
\label{eq30}
S(R)
\eq\f{R-C}{\sqrt{2\ve}}.
\eea
Then, the maximal Sharpe ratio in the range of $R\ge C$ is at 
$R^*=\arg\mathop{\max}_{R\ge C}S(R)=\f{V}{R_0-C}C^2+R_0$. 
Moreover, the maximum and minimum of the minimal investment risk per asset $\ve=\ve(R)$
are at 
$R_{\min}=\arg\mathop{\min}_{R\ge C} \ve(R)=R_0$
 and $R_{\max}=\arg\mathop{\max}_{R\ge C} \ve(R)=\infty$, respectively.
The squares of the Sharpe ratio are assessed as
\bea
S^2(R^*)
\eq
\f{\left\langle v^{-1}c^2\right\rangle}{\a-1}
\left(V+
\left(\f{\left\langle v^{-1}rc\right\rangle}
{\left\langle v^{-1}c^2\right\rangle}-1\right)^2
\right),\qquad\\
S^2(R_{\min})\eq
\f{\left\langle v^{-1}c^2\right\rangle}{\a-1}
\left(\f{\left\langle v^{-1}rc\right\rangle}
{\left\langle v^{-1}c^2\right\rangle}-1\right)^2,\\
S^2(R_{\max})\eq
\f{\left\langle v^{-1}c^2\right\rangle}{\a-1}V.
\eea
We obtain the following Pythagorean theorem of the Sharpe ratio:
\bea
S^2(R^*)\eq S^2(R_{\min})+S^2(R_{\max}).
\eea
Similar to what has been reported in the extant literature, 
this theorem is not dependent on $\a,C$ and the probabilities of $r_i,c_i,v_i$.
Further, the investment risk is summarized with respect to $C$: 
\bea
\ve\eq\f{\a-1}{2\left\langle v^{-1}r^2\right\rangle}
\left(R^2+\f{(C-C_0)^2}{V_r}\right),\\
C_0\eq R\f{\left\langle v^{-1}rc\right\rangle}
{\left\langle v^{-1}r^2\right\rangle},\\
V_r\eq
\f{\left\langle v^{-1}c^2\right\rangle}
{\left\langle v^{-1}r^2\right\rangle}-
\left(\f{\left\langle v^{-1}rc\right\rangle}
{\left\langle v^{-1}r^2\right\rangle}\right)^2,\\
S(C)\eq\f{R-C}{\sqrt{2\ve}}.
\eea
From this, the maximal Sharpe ratio in the range of $C\le R$ is at $C^*=\arg\mathop{\max}_{C\le R}S(C)=\f{V_r}{C_0-R}R^2+C_0$. Moreover, the maximum and minimum of the minimal investment risk per asset $\ve=\ve(C)$ are at 
$C_{\min}=\arg\mathop{\min}_{C\le R}\ve(C)=C_0$ and $C_{\max}=\arg\mathop{\max}_{C\le R}\ve(C)=-\infty$, respectively.
The squares of the Sharpe ratio are calculated as 
\bea
S^2(C^*)
\eq
\f{\left\langle v^{-1}r^2\right\rangle}{\a-1}
\left(V_r+
\left(\f{\left\langle v^{-1}rc\right\rangle}
{\left\langle v^{-1}r^2\right\rangle}-1\right)^2
\right),\qquad\\
S^2(C_{\min})\eq
\f{\left\langle v^{-1}r^2\right\rangle}{\a-1}
\left(\f{\left\langle v^{-1}rc\right\rangle}
{\left\langle v^{-1}r^2\right\rangle}-1\right)^2,\\
S^2(C_{\max})\eq
\f{\left\langle v^{-1}r^2\right\rangle}{\a-1}V_r.
\eea
We also obtain the following Pythagorean theorem of the Sharpe ratio: 
\bea
S^2(C^*)\eq S^2(C_{\min})+S^2(C_{\max}).
\eea

Next, let us compare the result of the annealed disordered investment system.
Applying a well-used analytical procedure of operations research,the minimal expected investment risk per asset $\ve_{\rm OR}=\lim_{N\to\infty}\f{1}{N}\mathop{\min}_{\vec{w}\in{\cal W}}E[{\cal H}(\vec{w})]$
is 
\bea
\ve_{\rm OR}\eq\f{\a}{2\left\langle v^{-1}c^2\right\rangle}
\left(
C^2+\f{\left(R-R_0
\right)^2}{V}
\right),
\eea
where $E[{\cal H}(\vec{w})]=\f{\a}{2}\sum_{i=1}^Nv_iw_i^2$ is used. From this, the proportion between the minimal expected investment risk per asset derived by  operations research 
$\ve_{\rm OR}$ and the minimal investment risk per asset $\ve$, the opportunity loss $\kappa=\f{\ve_{\rm OR}}{\ve}$, is solved as
\bea
\kappa\eq\f{\a}{\a-1}.
\eea
From this result, when $\a$ is close to 1, since 
the opportunity loss $\kappa$ is increasing, that is, since $\kappa=1$ is not satisfied, 
unfortunately, {the portfolio which can minimize the expected investment risk $E[{\cal H}(\vec{w})]$, 
$\vec{w}_{\rm OR}=\arg\mathop{\min}_{\vec{w}\in{\cal W}}E[{\cal H}(\vec{w})]$,
cannot minimize the investment risk ${\cal H}(\vec{w})$}.
Moreover, it transpires that the opportunity loss $\kappa$
is not dependent on $R,C$ and 
the probabilities of $r_i,c_i,v_i$.

\section{Numerical experiments\label{sec5}}
Here we focus on the case where the mean and square mean of the return $\bar{x}_{i\mu}$ are represented by 
$E[\bar{x}_{i\mu}]=r_i$ and $E[\bar{x}_{i\mu}^2]=(h_i+1)r_i^2$, respectively.
From this, the variance of 
the modified return $x_{i\mu}=\bar{x}_{i\mu}-E[\bar{x}_{i\mu}]$ 
is described by $v_i=V[x_{i\mu}]=h_ir_i^2$. Moreover, we set 
the relation between the purchase cost per unit of asset $i$, 
$c_i$, and the mean return $r_i$, $c_i=r_iz_i$, where 
$h_i,z_i$ are independently distributed and non-negative.
Then we assume that 
$r_i,h_i$ are distributed by the bounded Pareto distribution. These density functions $f_r(r_i),f_h(h_i)$ are defined as 
\bea
f_r(r_i)
\eq
\left\{
\begin{array}{ll}
\f{1-c_r}{u_r^{1-c_r}-l_r^{1-c_r}}r_i^{-c_r}&l_r\le r_i\le u_r\\
0&\text{otherwise}
\end{array}
\right.,\\
f_h(h_i)
\eq
\left\{
\begin{array}{ll}
\f{1-c_h}{u_h^{1-c_h}-l_h^{1-c_h}}h_i^{-c_h}&l_h\le h_i\le u_h\\
0&\text{otherwise}
\end{array}
\right.,\qquad
\eea
where $c_r,c_h>0$ are exponentials of the bounded Pareto distributions.
Moreover, $z_i$ is distributed uniformly with 
$0\le z_i\le1$.

From this numerical setting, the analytical procedure in the numerical experiments is organized as follows:
\begin{description}
\item[Step 1]Assign randomly $r_i,h_i$ with the bounded Pareto distributions
and evaluate the variance $v_i(=h_ir_i^2)$. Moreover, 
using $z_i$, which is distributed uniformly with $0\le z_i\le1$,
$c_i=r_iz_i$.
\item[Step 2]Assign the return $\bar{x}_{i\mu}$ with 
the Gaussian distribution $N(r_i,v_i)$, then the modified return 
$x_{i\mu}=\bar{x}_{i\mu}-E[\bar{x}_{i\mu}]$ is assessed. Moreover,
return matrix 
$X=\left\{\f{x_{i\mu}}{\sqrt{N}}\right\}\in{\bf R}^{N\times p}$ is set.
\item[Step 3]
Solve Wishart matrix $J=XX^{\rm T}\in{\bf R}^{N\times N}$ and its inverse matrix $J^{-1}$.
\item[Step 4]$\vec{c}^{\rm T}J^{-1}\vec{c},
\vec{c}^{\rm T}J^{-1}\vec{r},
\vec{r}^{\rm T}J^{-1}\vec{r}
$ are calculated.
\item[Step 5]
Using \sikis{eq8}{eq30}, we assess $\ve$ and $S$.
\end{description}

\begin{figure}[t] 
\begin{center}
\includegraphics[width=1.0\hsize]{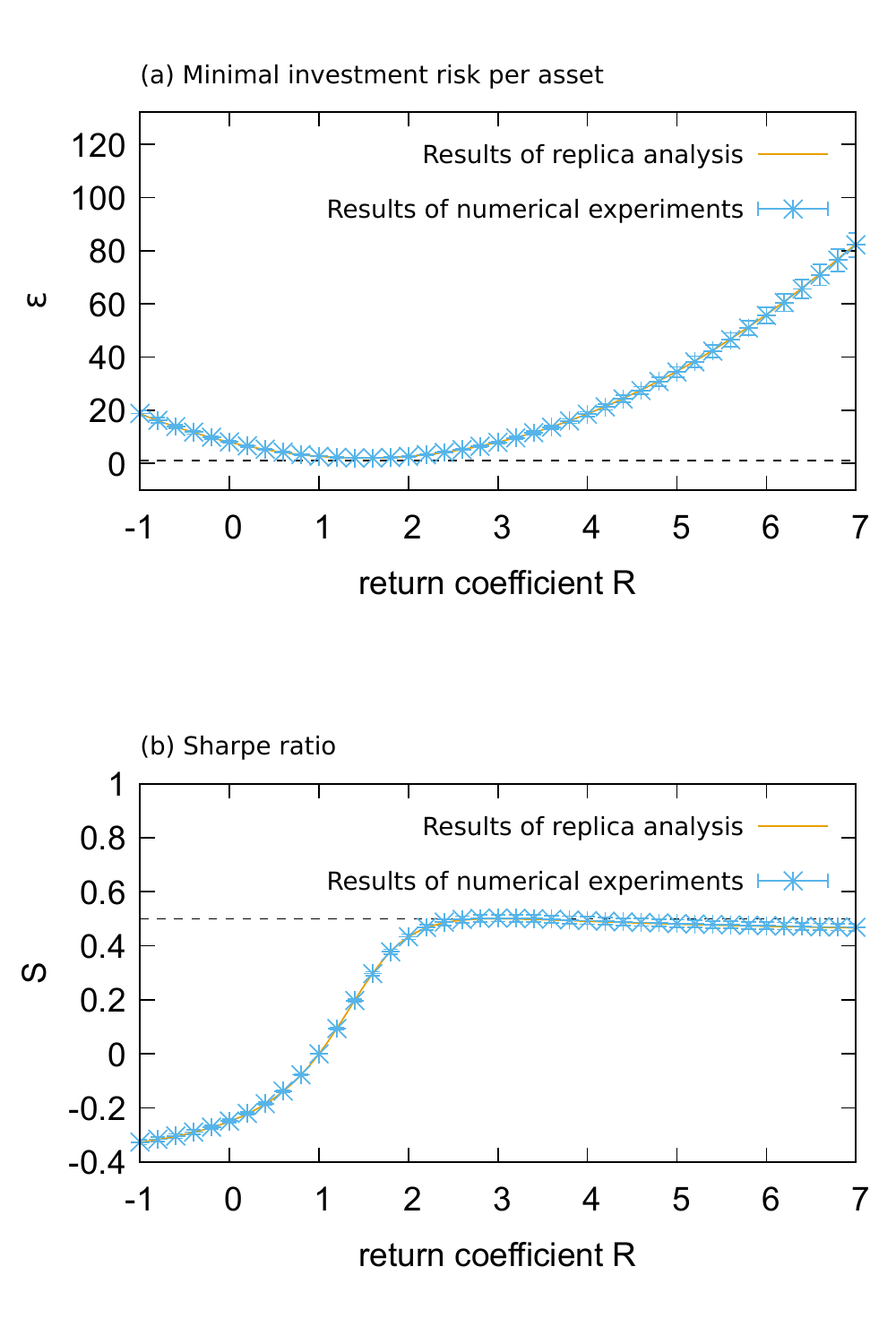}
\caption{
\label{Fig1}
Comparison of results from replica analysis and numerical experiments.
}
\end{center}
\end{figure}

Setting $N=1000,p=2000,(\a=p/N=2),C=1,l_r=l_h=1,u_r=u_h=2,c_r=c_h=2$,
we average the minimal investment risk per asset and 
the Sharpe ratio over 100 trials and compare 
it with results based on replica analysis. 
Fig. \ref{Fig1}(a) represents the return coefficient $R$ and 
the minimal investment risk $\ve$. 
Fig. \ref{Fig1}(b) represents the return coefficient $R$ and Sharpe ratio $S$.
The markers with error bars are the result of numerical experiments and the solid line is
the result of replica analysis.
The dotted line in Fig. \ref{Fig1}(a) is the minimum of the minimal investment risk 
$\ve(R_{\min})$ and 
the dotted line in Fig. \ref{Fig1}(b) is the maximum of Sharpe ratio $S(R^*)$.
From both figures, it is concluded that  
the results of replica analysis 
and the numerical experiments are consistent.

\section{Conclusion\label{sec6}}
We have improved on the analytical methods in the extant literature and discussed the investment risk minimization problem imposing cost and return constraints. In the budget constraint used in previous studies, the purchase cost per unit of each asset is not considered in detail. The investment risk minimization problem in previous studies has focused on considering the investment ratio as a portfolio (or rendering the purchase cost identical) without considering the purchase cost in actual investment market contexts. 
Because the optimal investment strategy varies depending on the size of working capital and the target figure, this study investigated the investment risk minimization problem imposing the cost constraint at the initial investment period and the return constraint at the final investment period by using replica analysis, with consideration of purchase cost, initial cost, and final return. The results suggest that the minimal investment risk per asset can be expressed as a function of the initial cost and final return. We compared the minimal investment risk derived by our proposed method with the minimal expected investment risk derived by an analytical method common to operations research and confirmed that the minimal investment risk is always lower than the minimal expected investment risk. We succeeded in deriving the opportunity loss from both results. We also confirmed that the Pythagorean theorem of the Sharpe ratio holds given the relationship between the maximum value of the Sharpe ratio corresponding to the minimum and maximum values of the minimal investment risk for the revenue coefficient and the cost coefficient. Finally, we show that the results derived by our proposed method are consistent with results from numerical experiments.

The Pythagorean theorem of the Sharpe ratio and the opportunity loss are macroscopic relations that do not depend on the distribution according to purchase cost or final return. It would be fruitful for future research to explore the generality of this finding. It would also be useful to investigate whether other macroscopic relations  hold in comparable contexts.

\section*{Acknowledgements}
The author is grateful for discussions with D. Tada and  I. Suzuki. 
This work was partially supported by
Grants-in-Aid Nos. 15K20999, 17K01260, and 17K01249; Research Project of the Institute of Economic Research Foundation at Kyoto University; and Research Project
No. 4 of the Kampo Foundation.


\bibliographystyle{jpsj}
\bibliography{sample20190105}

\begin{thebibliography}{10}

\bibitem{Luenberger1998InvestmentScience}
D.~G. Luenberger: {\em {Investment Science}} (Oxford University Press, 1998).

\bibitem{marcus2014investments}
Z.~Bodie, A.~Kane, and A.~Marcus: {\em Investments} (McGraw-Hill Education,
  2014).

\bibitem{1742-5468-2017-12-123402}
I.~Kondor, G.~Papp, and F.~Caccioli: Journal of Statistical Mechanics: Theory
  and Experiment {\bfseries 2017} (2017) 123402.

\bibitem{1742-5468-2016-12-123404}
I.~Varga-Haszonits, F.~Caccioli, and I.~Kondor: Journal of Statistical
  Mechanics: Theory and Experiment {\bfseries 2016} (2016) 123404.

\bibitem{doi:10.1080/1351847X.2011.601661}
F.~Caccioli, S.~Still, M.~Marsili, and I.~Kondor: The European Journal of
  Finance {\bfseries 19} (2013) 554.

\bibitem{KONDOR20071545}
I.~Kondor, S.~Pafka, and G.~Nagy: Journal of Banking \& Finance {\bfseries 31}
  (2007) 1545.

\bibitem{PAFKA2003487}
S.~Pafka and I.~Kondor: Physica A: Statistical Mechanics and its Applications
  {\bfseries 319} (2003) 487.

\bibitem{Pafka2002}
S.~Pafka and I.~Kondor: The European Physical Journal B - Condensed Matter and
  Complex Systems {\bfseries 27} (2002) 277.

\bibitem{Ciliberti2007}
S.~Ciliberti and M.~M$\acute{\rm e}$zard: The European Physical Journal B
  {\bfseries 57} (2007) 175.

\bibitem{doi:10.1080/14697680701422089}
S.~Ciliberti, I.~Kondor, and M.~M$\acute{\rm e}$zard: Quantitative Finance
  {\bfseries 7} (2007) 389.

\bibitem{110008689817}
T.~Shinzato: IEICE technical report {\bfseries 110} (2011) 23.

\bibitem{doi:10.7566/JPSJ.86.063802}
T.~Shinzato: Journal of the Physical Society of Japan {\bfseries 86} (2017)
  063802.

\bibitem{doi:10.7566/JPSJ.86.124804}
D.~Tada, H.~Yamamoto, and T.~Shinzato: Journal of the Physical Society of Japan
  {\bfseries 86} (2017) 124804.

\bibitem{SHINZATO2018986}
T.~Shinzato: Physica A: Statistical Mechanics and its Applications {\bfseries
  490} (2018) 986.

\bibitem{PhysRevE.94.052307}
T.~Shinzato: Phys. Rev. E {\bfseries 94} (2016) 052307.

\bibitem{PhysRevE.94.062102b}
T.~Shinzato: Phys. Rev. E {\bfseries 94} (2016) 062102.

\bibitem{10.1371/journal.pone.0134968}
T.~Shinzato and M.~Yasuda: PLOS ONE {\bfseries 10} (2015) e0134968.

\bibitem{10.1371/journal.pone.0133846}
T.~Shinzato: PLOS ONE {\bfseries 10} (2015) e0133846.

\bibitem{1742-5468-2017-2-023301}
T.~Shinzato: Journal of Statistical Mechanics: Theory and Experiment {\bfseries
  2017} (2017) 023301.

\bibitem{2018arXiv181006366S}
T.~{Shinzato}: arXiv e-prints  (2018) arXiv:1810.06366.

\bibitem{doi:10.7566/JPSJ.87.064801}
T.~Shinzato: Journal of the Physical Society of Japan {\bfseries 87} (2018)
  064801.

\bibitem{1742-5468-2018-2-023401}
T.~Shinzato: Journal of Statistical Mechanics: Theory and Experiment {\bfseries
  2018} (2018) 023401.

\bibitem{Ryosuke-Wakai2014}
R.~Wakai, T.~Shinzato, and Y.~Shimazaki: Journal of Japan Industrial Management
  Association {\bfseries 65} (2014) 17.

\end{thebibliography}

\end{document}